\documentclass[runningheads]{llncs}
\usepackage[T1]{fontenc}
\usepackage{graphicx}
\usepackage{amsmath}
\begin{document}
\title{“Do You Trust It… or Check It?”:
Exploring Children’s Verification Behaviors in Game-Based Human–AI Collaboration}
%
%
\author{Shuning Liu\inst{1} \and
Yuxin Song\inst{1} \and
Hanbing Li\inst{1} \and
Ziyue Wang\inst{2} \and
Mingying Lan\inst{1}\and
Fei Lyu\inst{1}\thanks{Corresponding author.}} 
\authorrunning{S. Liu et al.}
%
\institute{ Beijing University of Posts and Telecommunications, Beijing, China\\ \email{FeiLyu@bupt.edu.cn} \and Communication University of China, Beijing, China }
\maketitle              
\begin{abstract}
Children are increasingly interacting with AI agents in learning and problem-solving contexts, where AI suggestions may appear fluent, helpful, and socially engaging. However, children may not always be able to judge whether such suggestions are accurate, incomplete, or misleading. This paper presents a game-based human–AI collaboration probe designed to examine how children verify AI suggestions during task-based interaction. The system embeds AI suggestions within a biological knowledge reasoning game and compares three Agent interaction styles: partner-like, tool-like, and dual-state Agent. Instead of treating trust as a static post-task attitude, this study focuses on children’s situated verification behaviors, including acceptance, evidence checking, questioning, revision, and continued AI use. A preliminary evaluation with children aged 8–10 combined interaction logs, post-task ratings, and semi-structured interviews. The findings suggest that Agent style shaped children’s engagement and response pathways: partner-like interaction tended to support direct questioning, while tool-like interaction was more associated with evidence seeking. Children generally viewed AI as useful but fallible, and their questioning behaviors varied in reasoning depth. The study contributes design insights for supporting children’s critical, reflective, and evidence-based use of AI in collaborative tasks.

\end{abstract}
\section{Introduction}
Generative AI systems are increasingly entering children’s everyday learning, entertainment, and problem-solving activities. Through natural language dialogue, explanations, and socially expressive responses, AI agents may appear not only as information tools, but also as assistants or collaborative partners. For children, this creates new opportunities for inquiry-based learning and human–AI collaboration. However, it also raises important concerns: children may not always be able to judge whether AI suggestions are accurate, incomplete, or misleading, especially when the AI responds fluently and confidently.

Existing studies have examined children’s understanding of AI, their mental models of AI systems, and the role of AI literacy in helping children understand AI capabilities and limitations. Research on human–AI trust has also emphasized that appropriate reliance requires users to adopt AI suggestions when they are useful, while checking or questioning them when they may be wrong. However, less attention has been paid to how children actually respond to AI suggestions during concrete task-based collaboration. In particular, it remains unclear how children accept, question, verify, or revise AI suggestions when the AI provides incomplete or partially incorrect information.

To address this gap, this study presents a game-based human–AI collaboration probe designed for children aged 8–10. The task centers on biological knowledge reasoning and embeds AI suggestions within a structured game workflow. Instead of treating AI trust only as a post-task attitude, we examine children’s trust-related behaviors during interaction, including whether they continue the conversation, accept AI suggestions, seek evidence, question the AI, or revise their judgments. The system implements three Agent interaction conditions: a partner-like Agent, a tool-like Agent, and a dual-state Agent that dynamically switches between the two styles based on observable interaction signals.

Through a preliminary evaluation with children, this study explores how Agent style shapes children’s engagement and verification pathways, how children reason when they question AI, and how they perceive AI’s role and reliability after the task. The findings suggest that children’s trust in AI is not a single stable attitude, but is expressed through situated behaviors such as checking, questioning, and evidence-based correction. This paper contributes a game-based probe for studying children’s AI verification behaviors and offers design implications for building AI systems that support critical, reflective, and age-appropriate human–AI collaboration.

\section{Related Work}
\subsection{Children’s AI Literacy and AI Understanding}

With the widespread application of generative artificial intelligence in daily life, children have begun to interact and collaborate with AI systems in various task scenarios~\cite{Han2024GenerativeAIElementary,Newman2024DarthVader}. Existing research has shown that children's understanding of AI is not stable and singular, but is influenced by the form of AI interaction, task scenario, and system expression. They may understand AI as different roles, such as tools, assistants, or collaborative partners~\cite{Stanja2025AnthropomorphicConceptions}. Andries and Robertson found that although children can distinguish between AI and humans when interacting with smart speakers, they may still form inaccurate understandings of the intelligence, emotions, and agency of voice assistants~\cite{Andries2023Alexa}. With the popularization of generative AI, children may also gradually form mental models of AI's working mechanisms and capability boundaries through interactions with visual, textual, and dialogue generation systems~\cite{Kosoy2024ChildrenMentalModels}.

This understanding of AI roles and capabilities is closely related to AI literacy. Long and Magerko define AI literacy as the ability to understand, use, evaluate, and critically reflect on AI technologies, emphasizing the importance of understanding AI’s capability boundaries, judging AI outputs, and identifying potential errors~\cite{Long2020AILiteracy}. For children, AI literacy involves not only learning how to use AI tools, but also understanding how AI systems generate results and critically evaluating their outputs~\cite{Jia2025HolisticAILiteracy}. Therefore, children’s understanding of AI should not be regarded only as conceptual knowledge, but also as a basis for how they judge, adopt, or question AI suggestions in specific tasks.

However, existing research has mainly focused on how children understand and use AI systems~\cite{Stanja2025AnthropomorphicConceptions,Jia2025TechnologiesChildrenAI}, with less attention to how children respond to AI suggestions during task-based problem solving. In particular, it remains underexplored how children show acceptance, questioning, and verification behaviors when AI suggestions are incomplete or only partially correct.

\subsection{Trust, Reliance, and Verification in Human–AI Collaboration}

In human--AI collaboration, trust is an important factor that shapes how users understand, evaluate, and use information and suggestions provided by AI systems~\cite{Jacovi2021FormalizingTrust,Liao2022ResponsibleTrust}. For children, whose cognitive development, prior experience, and critical judgment abilities are still developing, AI systems with social expression and natural language interaction may lead to inappropriate trust or over-reliance~\cite{Pitts2025HumanAITrustEducation}. Therefore, the goal of children--AI collaboration is not simply to increase children’s trust in AI, but to support calibrated trust: children should be able to adopt AI suggestions when they are useful, while also checking and questioning AI outputs when they may be incomplete or incorrect.

Recent work has increasingly conceptualized trust in AI as a dynamic process rather than a fixed attitude formed after a single interaction. Trust can be influenced by system performance, user experience, and interaction context, and may change during continuous use~\cite{Jacovi2021FormalizingTrust}. In human--AI collaboration, this dynamic trust is reflected not only in users’ subjective evaluations, but also in their reliance behaviors, such as whether they adopt AI suggestions, continue to seek AI assistance, or check AI outputs in uncertain situations~\cite{Bucinca2021TrustThink,Vasconcelos2023Explanations}. Trust-related states may also inform the adjustment of AI interaction strategies, including explanation, communication style, and intervention level~\cite{Liao2022ResponsibleTrust,Ehsan2021ExpandingExplainability}. However, existing research has mostly focused on adults or general human--AI collaboration scenarios, with less attention to how children calibrate trust in task-based AI collaboration. Therefore, this paper examines children’s trust not as a single post-task attitude score, but through their acceptance, evidence checking, rejection, and revision behaviors during the task~\cite{AILitFramework2025}..

\subsection{Explainability, Anthropomorphism, and Interaction Style}
Existing research suggests that both explainability and anthropomorphism can influence how users understand and use AI systems, but their effects are not always linear. Explanations may help users understand the basis of AI outputs and reduce over-reliance~\cite{Ehsan2021ExpandingExplainability,Vasconcelos2023Explanations}, but they may also create a false sense of certainty or fail to support appropriate reliance in some contexts~\cite{Schoeffer2024Explanations}. Similarly, anthropomorphic cues may increase perceived closeness and trust resilience in cognitive agents~\cite{DeVisser2016AlmostHuman}, but anthropomorphic descriptions do not necessarily increase trust and may shape users’ perceptions differently depending on context~\cite{Inie2024ProbabilisticAutomation,Shi2025AnthropomorphismTrust}. For children, anthropomorphic language may also make AI appear more human-like, more reliable, or more authoritative.

Therefore, this paper treats explainability and anthropomorphic language as design resources that shape agent interaction styles. We compare three agent design archetypes: an instrumental agent that emphasizes explanation and task-oriented support, a companion agent that emphasizes social interaction and encouraging language, and a dynamic-switching agent that explores the possibility of adapting between these two forms.

\section{Methods}
This study employs an exploratory research method based on a system prototype, aiming to examine how children rely on, question, verify, and revise AI suggestions in gamified problem-solving tasks. Instead of viewing children's trust in AI as a static attitude after the task, this study focuses more on children's behavioral performance during the collaborative process, especially whether they will directly adopt, check evidence, raise questions, or revise their judgments when faced with AI suggestions of different reliability, and whether this will affect their subsequent use of AI assistance.

\subsection{System and Game Overview}

This study designed a game-based human–AI collaboration probe for children aged 8–10, centered on biological knowledge judgment and clue-based reasoning tasks. In the task, children were required to read task information, observe relevant clues, and make evidence-based judgments. The AI Agent was embedded within a continuous task workflow rather than functioning as an open-ended chatbot. It provided suggestions, explanations, or prompts only at predefined decision points, supporting children’s judgment process and triggering their verification and revision of AI-provided information.

\begin{figure}[t]
\centering
\includegraphics[width=0.8\textwidth]{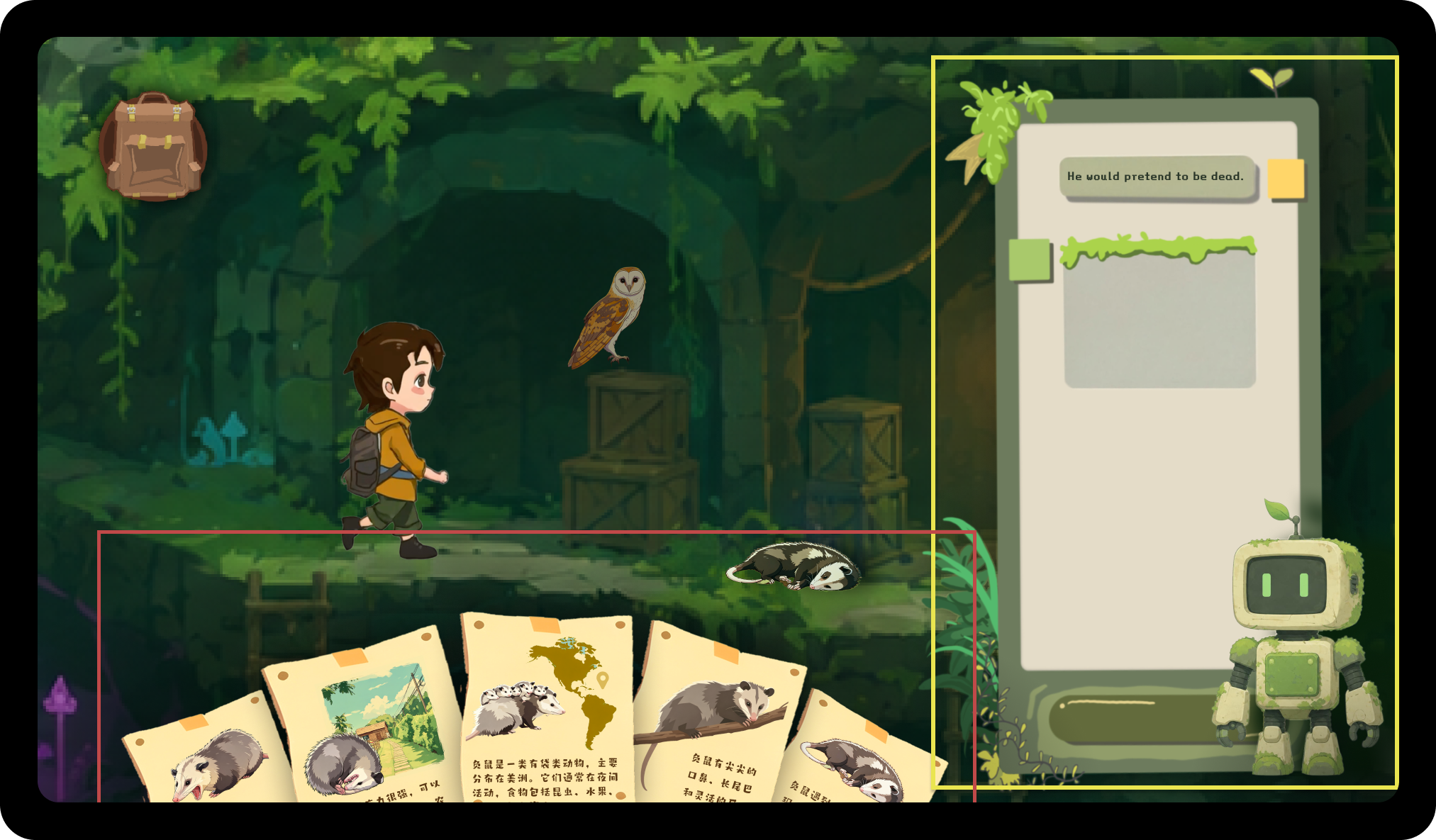}
\caption{Child-facing interface of the game-based collaboration task, including the task area, clue area, and Agent feedback area.}
\label{fig:child_interface}
\end{figure}

The child-facing interface consisted of three main areas: the task area, the clue area, and the Agent dialogue area (Fig. 1). The task area presented the current problem context, the clue area provided visual or textual evidence relevant to the judgment, and the Agent dialogue area displayed the AI’s suggestions, explanations, or prompts while supporting the recording of children’s verification choices and final answers.

\begin{figure}[htbp]
\centering
\includegraphics[width=0.88\textwidth]{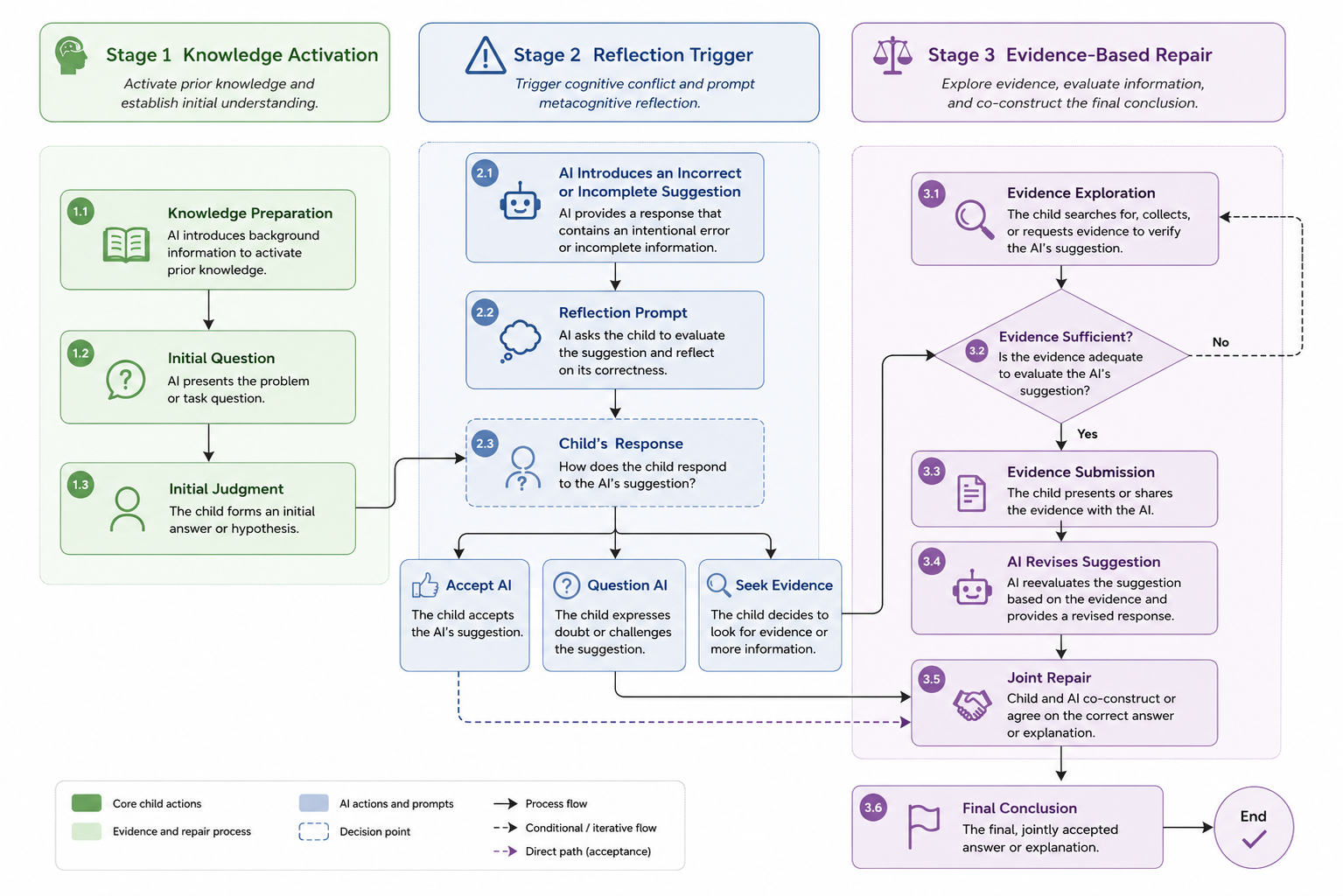}
\caption{Branching task workflow with three phases of knowledge activation, reflection triggering, and evidence-based repair.}
\label{fig:workflow}
\end{figure}

Each key decision point followed the same workflow. First, children read a knowledge card and formed an initial judgment before the AI intervention. They then received an incomplete or partially inaccurate suggestion from the AI Agent and chose whether to accept it, question it, or seek additional evidence. If children entered the evidence exploration process, they were asked to submit evidence based on the provided clues and collaboratively revise the suggestion with the AI to reach a final conclusion. Through this workflow, the study observed whether children treated AI suggestions as direct answers or as information sources that required verification through evidence checking, questioning, and judgment revision.

\begin{figure}[t]
\centering
\includegraphics[width=0.88\textwidth]{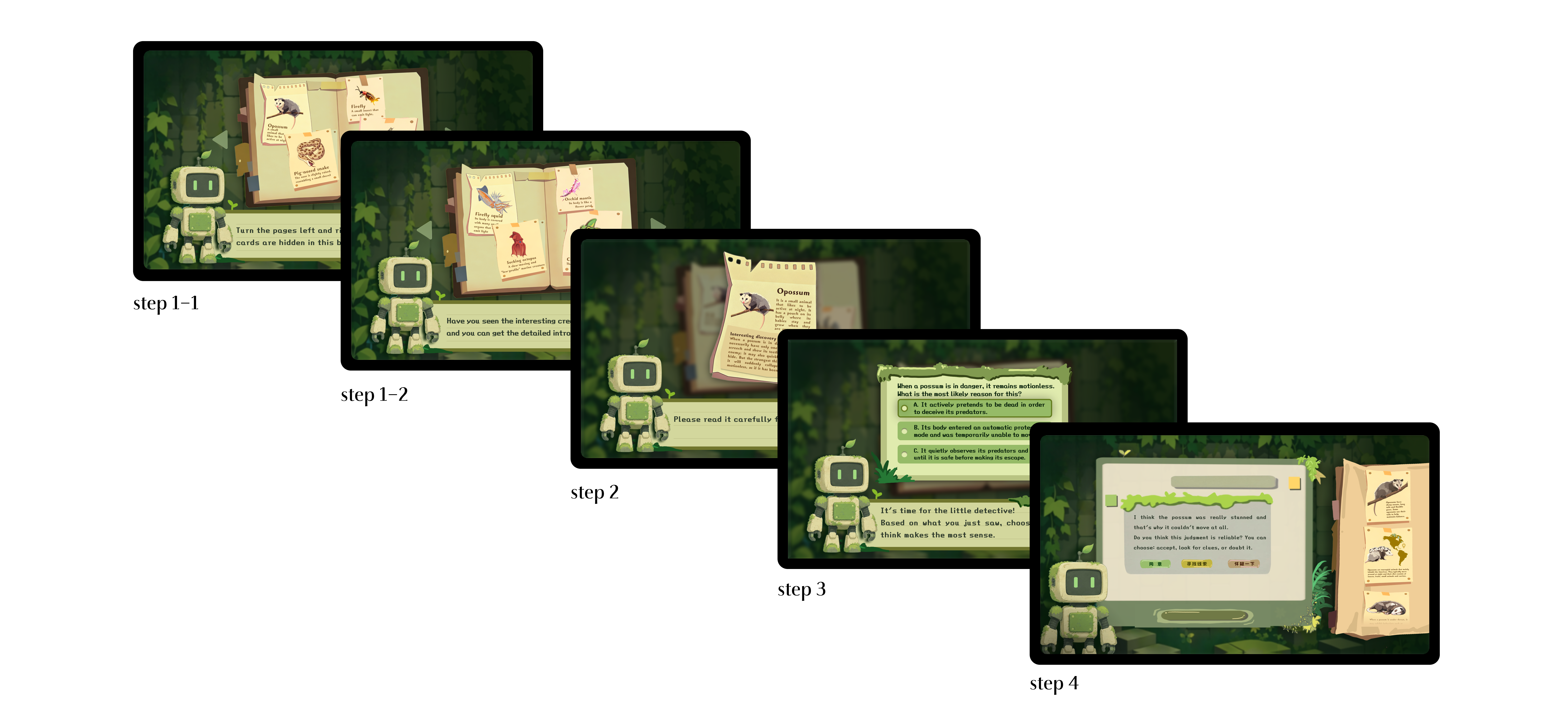}
\caption{A key game-level example showing how a partially correct AI suggestion is embedded in a clue-based reasoning task.}
\label{fig:key_level_example}
\end{figure}

\subsection{Agent Interaction Conditions }
This study implemented three Agent interaction conditions: partner-like Agent, tool-like Agent, and dual-state Agent. Across conditions, the core AI suggestions remained identical and partially correct; the conditions differed only in language style, explanation format, and interaction prompts.

\subsubsection{Partner-like Agent} 
The Agent emphasized social companionship and joint exploration. It used friendlier, more encouraging, and collaborative expressions, such as “Let’s take a look together” or “I think this clue might be important.” 

\subsubsection{Tool-like Agent} 
The Agent emphasized knowledge support and task assistance. Its language reduced subjective, emotional, and role-based expressions, avoiding human-like phrases such as “That’s amazing” or “You reminded me.” 

\subsubsection{Dual-state Agent} The dynamically switching Agent adjusted between partner-like and tool-like expressions. The system did not claim to accurately measure children’s internal trust states; instead, it used observable behaviors as indicators of reliance. For example, during general chat, the Agent adopted a more partner-like style to help children enter the task naturally. When children expressed doubt or searched for evidence, the Agent shifted toward a tool-like response. 

\subsection{System Architecture}
The system architecture consists of a task knowledge base, a response control module, a style-switching module, and a logging module. The front end uses Unity to build the game scene and dialogue interface, and communicates with the back end through HTTP requests. The back end uses FastAPI to provide the Agent service interface, and the response generation model is qwen-plus.

\subsubsection{Task RAG} 
The task knowledge base stores biological knowledge, key phenomena, and controllable error points used in the game levels. The tool-like Agent retrieves relevant knowledge fragments according to the child's input and generates responses based on the retrieved content. Knowledge retrieval is implemented through vector search, using text-embedding-v4 as the embedding model, in order to improve the accuracy of knowledge matching in open-ended dialogue. This module ensures that the content generated by the tool-like Agent is grounded in task-related knowledge and remains explainable.

\subsubsection{Response Control Module} 
The module manages the task flow and dialogue states. A state machine is used to determine whether the system is currently in free chat, error-triggering, reflection selection, AI revision, or final conclusion stages. Based on the current state, the module decides whether to return general conversational content or controlled error content. This module ensures that the dialogue does not deviate from the predefined learning task.

\subsubsection{Style-Switching Module} 
This module controls the dynamic transition of the dual-state Agent between the partner-like state and the tool-like state. The system calculates a trust score based on children's real-time interaction signals, and uses the change in trust to determine which type of Agent response should be returned in the current interaction round.

The trust computation consists of two steps. First, the system calculates an immediate trust score based on the child's textual response, choice behavior, interaction behavior, and response time in the $t$-th interaction round:

\begin{equation}
S_t = a \cdot Text_t + b \cdot Choice_t + c \cdot Behavior_t + d \cdot Time_t
\end{equation}

where $S_t$ denotes the immediate trust score in the $t$-th interaction round. $Text_t$ represents whether the child's textual response indicates acceptance, questioning, or a request for evidence. $Choice_t$ represents the child's selection among options such as ``accept'', ``seek evidence'', or ``question''. $Behavior_t$ captures whether the child actively provides reasons, relies on the AI's answer, or continues to interact with the task. $Time_t$ represents the response time interval. The weights satisfy:

\begin{equation}
a + b + c + d = 1
\end{equation}

The system then updates the overall trust score by combining the immediate trust score with the trust score from the previous interaction round:

\begin{equation}
T_t = \lambda \cdot T_{t-1} + (1-\lambda) \cdot S_t
\end{equation}

where $T_t$ denotes the overall trust score after the $t$-th interaction round, $T_{t-1}$ denotes the trust score from the previous round, and $\lambda$ is a smoothing coefficient. This update mechanism allows the trust state to be influenced by both the current interaction and previous interaction history, while avoiding frequent Agent switching within a short period of time.

Finally, the system determines the Agent style according to the current trust score. When the trust score indicates over-reliance, the Agent shifts toward a more tool-like state by providing more explicit evidence and explanation. When the trust score indicates low engagement or insufficient reliance, the Agent shifts toward a more partner-like state by using more encouraging and socially supportive responses.

\subsubsection{User Behavior Logging} 
The module records children’s input, Agent responses, state transitions, choice results, and retrieved information in each conversation. The system generates an independent log file for each session, enabling subsequent analysis of how children accept, question, and revise AI suggestions.

\subsection{Study Procedure}
This study adopted a between-group experimental design and recruited a total of 12 children aged 8–10 years. Participants were assigned to one of three Agent conditions, with 4 children in each condition. The between-group design was adopted to reduce the learning effect and direct comparison effect among children across different Agent styles.

The experimental process consists of four stages. First, the researcher introduces the task rules and interface operations to the children and confirms their understanding of the basic operation methods. Second, the children enter the Gamification task stage. This stage revolves around the same biological knowledge point. First, through the knowledge judgment task, it examines the children's process of verifying and correcting partially correct AI suggestions, and then through the contextualized application task, it further observes their subsequent seeking and adoption of the same agent assistance. Third, after the task is completed, the children complete a brief AI reliability and trust rating. Fourth, the researcher conducts a semi-structured interview to ask the children why they believe, question, or correct AI suggestions, as well as how they understand the role, reliability, and capability boundaries of AI.

Since the task includes intentionally designed partially correct AI suggestions, after the experiment, researchers will explain to the children that AI may provide incomplete or suggestions that need to be checked, and emphasize that it is reasonable and important to review evidence and maintain one's own judgment when using AI.

\section{Findings}

This study conducted a preliminary experimental evaluation. We analyzed three types of data from the prototype evaluation: interaction logs, post-task ratings, and semi-structured interviews. Interaction logs were coded for children’s response pathways, post-task ratings were analyzed using descriptive statistics, and interviews were coded thematically to interpret children’s understanding of AI roles, reliability, and judgment strategies.

\subsection{Agent Style Shaped Engagement and Response Entry}
Behavioral logs showed that Agent style influenced children’s engagement before they made explicit trust decisions. Among the 11 valid sessions, 10 children entered the reflection node, where the AI introduced an incorrect explanation and asked children to decide whether to accept, seek evidence, or question the suggestion. All children in the partner-like condition (A) and the dual-state (AB) entered this node, while one child in the tool-like condition (B) stopped during free chat. Since all conditions used the same task-triggering mechanism, this suggests that Agent expression may affect children’s willingness to continue interacting and reach the reflection process.

Children’s response tendencies differed across Agent conditions. In the partner-like condition, all four children chose to question the AI, suggesting that socially expressive interaction may encourage direct disagreement. In the tool-like condition, responses were more distributed: two children chose to seek evidence, one questioned the AI, and two did not complete a response choice. The dynamic-switching condition showed a mixed pattern, with one child seeking evidence and one accepting the AI suggestion. Both valid AB sessions started with a companion-like entry and shifted toward a tool-like response during reflection or repair, suggesting the feasibility of the routing mechanism, although the small sample size does not support claims of a stable behavioral effect.
\begin{table}[t]
\centering
\small
\caption{Behavioral pathways across Agent conditions.}
\label{tab:behavioral_pathways}
\begin{tabular}{lcccccccc}
\hline
\textbf{Condition} & \textbf{N} & \textbf{Reflection} & \textbf{Early stop} & \textbf{Accept} & \textbf{Seek} & \textbf{Question} & \textbf{Repair} & \textbf{Success} \\
\hline
A & 4 & 4 & 0 & 0 & 0 & 4 & 4 & 2 \\
B & 5 & 4 & 1 & 0 & 2 & 1 & 3 & 4 \\
AB & 2 & 2 & 0 & 1 & 1 & 0 & 2 & 1 \\
\hline
\end{tabular}
\end{table}

\subsection{Children’s Verification Behaviors Varied in Pathways and Reasoning Depth}
Children’s responses to the AI suggestion were not limited to simple acceptance or rejection. Across the 11 valid sessions, only one child accepted the AI suggestion, while three chose to seek evidence and five questioned the AI. Nine sessions reached the AI repair state, indicating that most children moved from the initial AI suggestion into a process of verification or correction. These results suggest that children’s trust in AI was expressed through situated behaviors such as checking, questioning, and revision, rather than through a single decision to trust or distrust the AI.

However, children’s questioning varied in reasoning depth. Among the children who questioned the AI, some expressed intuitive disagreement, while others provided conceptual corrections or referred to task evidence such as images or knowledge cards. This suggests that questioning AI should not be treated as a uniform indicator of critical AI use; the quality of children’s reasoning and evidence use is also important. In the second task, only one child triggered robot assistance, so continued AI use was treated as a supplementary indicator rather than a full measure of collaboration quality.

\begin{table}[t]
\centering
\small
\caption{Coding of children's reasons for questioning the AI.}
\label{tab:reason_quality}
\begin{tabular}{p{3.8cm}p{6.2cm}c}
\hline
\textbf{Reason type} & \textbf{Description} & \textbf{Count} \\
\hline
Intuitive disagreement & The child expressed doubt without clear evidence. & 1 \\
Conceptual correction & The child identified the conceptual problem in the AI suggestion. & 3 \\
Evidence-based correction & The child referred to task evidence such as images or knowledge cards. & 1 \\
\hline
\end{tabular}
\end{table}

\subsection{Children Viewed AI as Useful but Fallible}
Post-task ratings showed that children generally viewed the AI as useful but fallible. Across the three Agent conditions, AI trust and reliance remained at a moderate level ($M=3.33$), while error awareness was consistently high ($M=4.83$), suggesting that children did not treat the AI as an always-correct source. Children also rated AI explanations as helpful ($M=4.50$), indicating that explanations were perceived as useful task support. However, their role perception remained largely functional: tool perception ($M=4.67$) was higher than partner perception ($M=3.33$). Although the dynamic-switching condition showed relatively higher partner perception ($M=4.25$), children in this condition still rated the AI strongly as a tool ($M=5.00$), suggesting that the AI was primarily understood as a system for analyzing information and supporting decisions rather than as a fully collaborative partner.

\begin{table}[t]
\centering
\small
\caption{Descriptive post-task rating results across Agent conditions.}
\label{tab:post_task_ratings}
\begin{tabular}{p{4.2cm}cccc}
\hline
\textbf{Dimension} & \textbf{A} & \textbf{B} & \textbf{AB} & \textbf{Overall M} \\
\hline
AI trust and reliance & 3.25 & 3.25 & 3.50 & 3.33 \\
Error awareness & 5.00 & 4.75 & 4.75 & 4.83 \\
Explanation usefulness & 4.25 & 4.50 & 4.75 & 4.50 \\
Collaboration perception & 2.50 & 3.50 & 1.75 & 2.58 \\
Tool perception & 4.50 & 4.50 & 5.00 & 4.67 \\
Partner perception & 3.00 & 2.75 & 4.25 & 3.33 \\
Platform usability & 4.00 & 4.08 & 4.12 & 4.07 \\
\hline
\end{tabular}
\end{table}

The platform usability ratings were generally positive, indicating that children could understand and operate the game system. However, open-ended comments pointed to several design issues, including visual clarity, card readability, response speed, task difficulty, and the naturalness of AI interaction. These results suggest that children’s collaboration experience was shaped not only by Agent style, but also by the broader quality of the interface, task flow, and system fluency.

\section{Discussion}

The findings suggest that Agent style should be designed as task-based scaffolding rather than a fixed personality. The companion-like Agent supported engagement and direct questioning, while the tool-like Agent was more related to evidence-seeking behaviors. Therefore, children’s AI systems may benefit from combining both styles: using social expression to sustain participation, and shifting toward structured, evidence-based explanations when children need to evaluate or revise AI suggestions.

The findings also suggest that AI literacy should be cultivated through situated interaction, not only through abstract knowledge about AI. Children showed different forms of critical AI use, including seeking evidence, questioning AI, and correcting AI explanations, but their reasoning depth varied. Therefore, future AI literacy designs should help children practice how to judge AI outputs, compare them with evidence, explain their reasoning, and distinguish between their own judgments and AI-provided information.

This study has several limitations. First, the evaluation was conducted with a small number of participants, so the results should be interpreted as preliminary trends rather than stable causal conclusions. Future work should involve a larger and more balanced sample, with better control of children’s prior AI experience and task ability. Second, the current dynamic trust model mainly maps observable behaviors to trust states, but does not fully explain the underlying reasons for children’s trust changes. Children’s trust may be influenced by multiple factors, including task understanding, AI expression, cognitive load, and perceived AI authority.

Future research can further examine whether children’s critical AI use can transfer from structured game tasks to more open-ended scenarios, such as scientific inquiry, creative generation, or learning assistance. In addition, future systems could include more explicit reflection prompts that ask children to explain why they accept, question, or revise AI suggestions. This would help reveal how children distinguish between their own judgments and AI-provided information during multi-round human–AI collaboration.

\begin{credits}
\subsubsection{\ackname} 
The authors would like to thank Prof. Fei Lyu and Prof. Mingying Lan from ‌Beijing University of Posts and Telecommunications‌ for the valuable guidance and support. This work was supported by the National Natural Science Foundation of China (Grant No. 61972045).

\subsubsection{\discintname}
The authors have no competing interests to declare.
\end{credits}
%
%
%
\bibliographystyle{splncs04}
\bibliography{report}

\end{document}